# Atmospheric $O_2$ from astronomical data

A. Grieco[1], G. Candeo[2], P. Marziani[2]

[1]*Università degli Studi di Padova, Padova, Italy*
[2]*Osservatorio Astronomico di Padova, INAF, Padova, Italy*

**Abstract**

Environmental research aimed at monitoring and predicting $O_2$ depletion is still lacking or in need of improvement, in spite of many attempts to find a relation between atmospheric gas content and climate variability. The aim of the present project is to determine accurate historical sequences of the atmospheric $O_2$ depletion by using the telluric lines present in stellar spectra. A better understanding of the role of oxygen in atmospheric thermal equilibrium may become possible if high-resolution spectroscopic observations are carried out for different airmasses, in different seasons, for different places, and if variations are monitored year by year. The astronomical spectroscopic technique involves mainly the investigation of the absorption features in high-resolution stellar spectra, but we are also considering whether accurate measures of the atmospheric $O_2$ abundances can be obtained from medium and low resolution stellar spectra.

**Introduction: climate change and $O_2$ depletion**

The topic of global changes in climate is a relatively new field of study. Much environmental research is being conducted to find a relation between atmospheric gas content and climate variability. Oxygen seems to be one of the most significant keys to provide past climates. Many complex mechanisms work jointly together in balancing $O_2$ concentrations in the atmosphere (Lenton & Watson 2000), also involving P, N, C cycles. Model predictions indicate that global climate changes and pollution could lead to a depletion of the oxygen in many Earth systems. However, the environmental research aimed at monitoring and predicting $O_2$ depletion is lacking or needs to be improved (European Commission 2007). Theoretical models for better describing and predicting global climate variations depend strongly on the accuracy of measures of $O_2$ content in different Earth's ecosystem. $O_2$ values derived from standard methods display too large variability. The complex nature of spatial-temporal variation in atmospheric gas composition at small spatial scales may underlie the observed variability. Therefore, a new program for obtaining accurate measures of oxygen abundances and spatial distribution is highly relevant for climate and atmospheric energy balance studies.Atmospheric conditions have been always a fundamental issue for astronomers in the selection of astronomical observing sites and in the routine operation of an observing site. Nevertheless, despite the great potentialities of astronomical observing sites, astronomy has had a marginal role in this field of studies (Somov & Khlystov 1993). Night sky features in astronomical observations are almost always considered to be an inconvenience. Sky spectra obtained during typical observing nights cannot rival more focused observation obtained through dedicated instrumentation (see for example, CESAR; Compact Echelle Spetrograph for Aereonomic Research; Slanger et al.2007). Nonetheless, night sky spectra are plentiful and they are being collected in several places around the world at a very fast pace. They can provide information of the Earth's atmosphere $O_2$ for a time lapse that can be as long as 50 years if digitized photographic plates are considered.

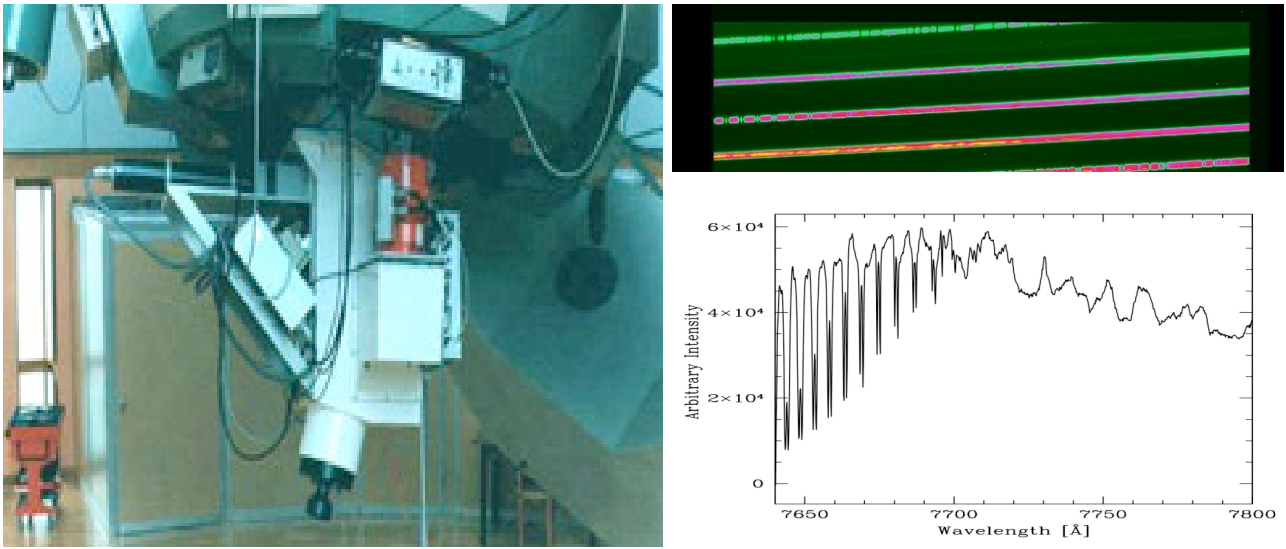

Figure 1 – Left: the Echelle spectrograph mounted at the Cassegrain focus of the 1.82 m telescope of the Padova Astronomical Observatory. Upper Right: CCD frame with part of the Echellegrams recorded in a single exposure. Lower Right: the spectrum of order #29, covering part of the A band.

## The Project

A purpose of the study we are beginning to articulate is the determination of an historical sequence of the atmospheric $O_2$ abundance, obtained from telluric lines present in stellar spectra. A second aim is to organize a network program to measure the $O_2$ atmospheric content and monitoring its spatial and temporal distribution. The telluric absorption lines in the solar or stellar spectra can provide precise measures of the concentration of different gaseous component in the Earth's atmosphere. Of special relevance are the lines of molecular oxygen, of $CO_2$, and the emission features of OH. Using astronomical spectroscopic procedures, accurate measurements of $O_2$ content can be derived from high resolution spectra. However, most of the stellar spectra present in astronomical archive are obtained with low and medium resolving power.

It is possible to resolve the A band structure with a typical echelle spectrograph. This is clearly seen in the spectrum of the Be star HD 149757 obtained with the Echelle spectrograph at Asiago Observatory at Mount Ekar (Figure 1), with a spectral resolution of approx. 0.3 Å FWHM (R~25000). Low resolution observations of the A and B band are routinely recorded in the red/near IR. The spectral resolution is however insufficient to distinguish individual features. Some features are still retained at resolution FWHM ~ 2 Å (middle panel), but they are almost fully lost if spectral resolution is ~1000 or below. This is visible in the bottom panel of the figure below, where the A band profile (as before, in the ideal case of infinite S/N) is plotted.

Therefore, efforts will be also focused at developing standardised methodology to obtain precise $O_2$ abundances from Fraunhofer A and B molecular oxygen band or from other different oxygen spectral features, using spectra of different dispersion (available at Asiago Observatory at Mount Ekar and other astrophysical sites).

## The A band

The $O_2$ A and B-band, respectively centered 13122 $cm^{-1}$ and 14550 $cm^{-1}$, are the most prominent near-infrared absorption features in the Earth's atmosphere. The A band has been extensively used in aereonomie to determine cloud top heights, cloud optical properties and for determination of atmospheric pressure at earth surface. The A band arises from the 0 – 0 vibration levels of the $b\ ^3\Sigma^+_g - X\ ^1\Sigma^+_g$ electronic transition of molecular oxygen. The structure of the A band from the HITRAN04 database assuming T = 296 °K (Rothman et al. 2005), is shown in the top panel of Figure 2 at very high resolution. A rough and quick measure of the column density of absorbing molecules can be obtained from the integrated band strength and from the equivalent width of the band, which is relatively straightforward to measure (of the order of 10 Å). More accurate information may however come from profile models of both the A and B band. Models of the observed A and B band emission could be constructed assuming standard (1976 US Standard Atmosphere) or *derived* physical conditions, and computing line-by-line radiation tranfer through the earth's atmosphere (see e.g. Werner et al. 2001), for example employing LBLRTM (Clough et al. 2005). Computations can be based on the HITRAN04 database (and developments; Rothman et al. 2005) or other extensive molecular databases.

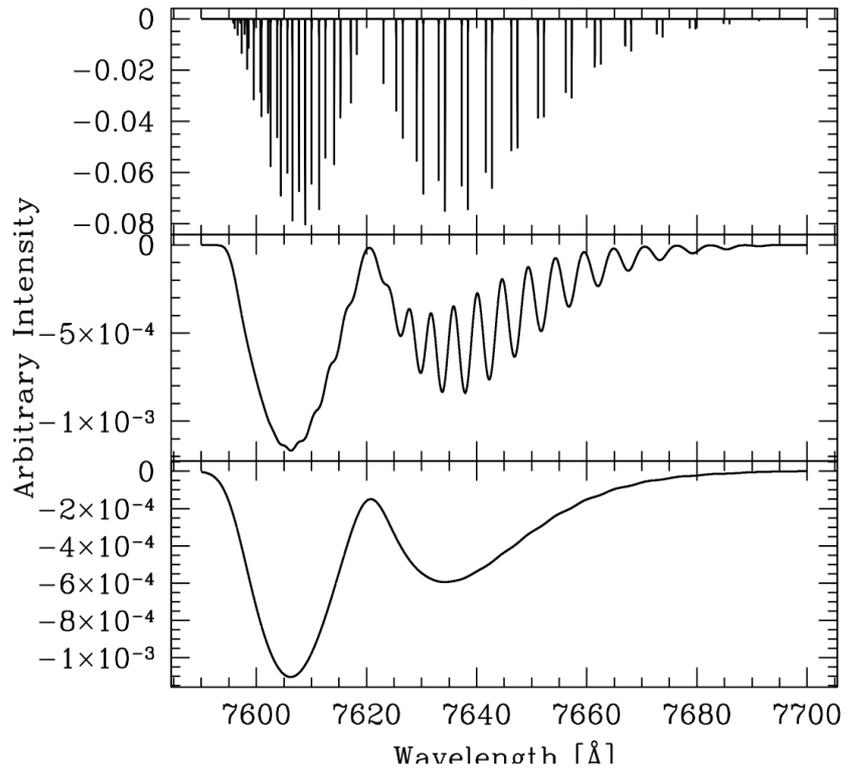

Figure 2 – Appearance of the A band with three different instrumental resolutions. The upper panel shows the fully resolved structure, while the bottom panel shows the blended features at R~1000.

**Methods for the determination of the atmospheric $O_2$ Abundance**

The possibility of using telluric lines to study global $O_2$ content and other atmospheric characteristics has been illustrated with solar observation carried out in Moscov in the visible and near infrared. In their work, Somov and Khlystov (1993) measured the equivalent widths of the $O_2$ spectral line λ 629.5 nm and $CO_2$ line λ 2070.0 nm present in the observation obtained at Moscow Observatory in 1992, and compare this spectra with archive equal observations made 20 years before. Using high resolution solar spectra Somov and Khlystov (1993) demonstrated that the atmospheric content of oxygen in the Moscow area did not change over two decades from 1970 to 1990 years (to within an error of 5%). On the contrary, during the same period, the content of carbon dioxide increased by 48%, lending credit to the most pessimistic forecasts about the dynamics of growth of the $CO_2$ content in the Earth's atmosphere. We can attempt to derive the atmospheric $O_2$ abundances from high resolution stellar spectra that are available at Asiago, Italy. Indeed, stellar spectra obtained with the Echelle Spectrograph at the Ekar 1.82m (Asiago, Italy) since 1993 to the present are available. The Echelle spectrograph at the Cassegrain focus of the 1.82 m telescope of the Asiago Observatory at Mount Ekar (displayed in Figure 1) provides high dispersion spectra [from 5 Å /mm in the far blue (3450 Å), to 10 Å/mm at 6600 Å]. At this resolving power, the methods of Vince et al., and Somov & Khlystov might become straightforwardly applicable to night-time observations of bright stars whereas the individual rotational lines of the A bands are resolved. Also ESO archive Coudé Echelle Spectra, with a spectral resolution up to 235,000 in the 346 - 1028 nm region will be considered.

The equivalent width (W) of the absorption lines (or the energy absorbed to the star's light by the atmospheric gas) depends on the column density (N) of the particular atmospheric component along the line of sight; it also varies with atmospheric mass (M) and air temperature (T), and other atomic variables. Seasonal and diurnal variations of the line equivalent width (due to variations in air density, temperature and photosynthesis) have to be evaluated. The equivalent width can be expressed as a function of the T and N along the line of sight: $W = (\alpha N/T) \exp(-hcF/kT)$, where F is the energy of the lower level of the line in cm$^{-1}$, $\alpha$, h, c and k are constant factors, $E_j = BhcJ(J + 1)$ is the energy of the lower molecular level, J is the rotational quantum number of the level j, and B is the rotation constant of the $O_2$ molecule. The method yields the total amount of the constituent across the whole height of the atmosphere, providing an efficient tool for studying global changes in the Earth's atmosphere composition. To determine whether this concentration varies in time, (W, M) plots are needed for different years. The predicted $W = f(M)$ are represented by curves varying linearly in the range of air mass between 1 and 2, but showing a vertical shifts M if the $O_2$ atmospheric concentration has changed during this period.

### Other Relevant Spectral Features Recorded in Astronomical Spectra

Low dispersion spectrographs which are part of the normal equipment of major observatories can record the stronget Meinel OH band up to 900 nm. The individual lines of the P and Q branches are resolved also in low-resolution spectra, with moderate resolution (~1000). The origin of OH in the upper atmosphere has been linked to reactions between hydrogen, ozone, and oxygen, namely: $H + O_3 \rightarrow OH + O_2$ and $OH + O \rightarrow O_2 + H + e$. The largest telescopes are now equipped with NIR spectrograph operating a relatively high resolution. This will make possible to retrieve information on the $CO_2$ feature at 1-2 mm, and hence on the $CO_2$ abundance.

### Conclusions

In summary a variety of astronomical observations can help constrain [$O_2$], [$O_3$], [$CO_2$] in the past decades. Archived astronomical observations, even if not dedicated, can provide an historical database that can be used for comparison with new, more accurate diurnal data that will be provided by space borne or ground based instrumentation. Much work remains to be done to assess the accuracy of any [$O_2$] determination based on low-resolution spectra.